\documentclass[aps,prd,twocolumn,showpacs,showkeys,amsmath,amssymb,
groupedaddress]{revtex4}

\usepackage{times}

\newcommand{\prt}{\partial}
\newcommand{\prm}{\prime}
\newcommand{\mrm}{\mathrm} 

\begin{document}

\preprint{}

\title{Perturbations on steady spherical accretion in 
Schwarzschild geometry}

\author{Tapan Naskar$^1$}
\email{tptn@mahendra.iacs.res.in}
\affiliation{$^1$Department of Theoretical Physics \\
Indian Association for the Cultivation of Science \\
Jadavpur, Kolkata 700032, India}
%\altaffiliation[Also at: ]{Inter--University Centre for Astronomy 
%and Astrophysics, 
%Post Bag 4, Ganeshkhind, Pune University Campus, Pune 411007, India}

\author{Nabajit Chakravarty$^2$}
\email{tpnc@mahendra.iacs.res.in}
\affiliation{$^3$Department of Theoretical Physics \\
Indian Association for the Cultivation of Science \\
Jadavpur, Kolkata 700032, India}
\altaffiliation[Also at: ]{Positional Astronomy Centre\\
P-546, Block 'N' New Alipure, Kolkata 700053, India}

\author{Jayanta K. Bhattacharjee$^3$}
\email{tpjkb@mahendra.iacs.res.in}
\affiliation{$^3$Department of Theoretical Physics \\
Indian Association for the Cultivation of Science \\
Jadavpur, Kolkata 700032, India}

\author{Arnab K. Ray$^4$}
\email{akr@iucaa.ernet.in}
\affiliation{$^4$Inter--University Centre for Astronomy and Astrophysics \\
Post Bag 4, Ganeshkhind, Pune University Campus, Pune 411007, India}

\date{\today}

\begin{abstract}
The stationary background flow in the spherically symmetric infall
of a compressible fluid, coupled to the space-time defined by the
static Schwarzschild metric, has been subjected to linearized
perturbations. The perturbative procedure is based on the continuity 
condition and it shows that the coupling of the flow with the geometry 
of space-time brings about greater stability for the flow, to the
extent that the amplitude of the perturbation, treated as a standing
wave, decays in time, as opposed to the amplitude remaining constant
in the Newtonian limit. In qualitative terms this situation simulates
the effect of a dissipative mechanism in the classical Bondi accretion
flow, defined in the Newtonian construct of space and time. As a
result of this approach it becomes impossible to define an acoustic 
metric for a
conserved spherically symmetric flow, described within the framework
of Schwarzschild geometry. In keeping with this view, the
perturbation, considered separately as a high-frequency travelling
wave, also has its amplitude reduced. 
\end{abstract}


\maketitle

\section{Introduction}
\label{sec1}

The spherically symmetric model of astrophysical accretion continues
to enjoy an abiding appeal among researchers in accretion astrophysics, 
starting with the seminal paper written by Bondi~\cite{bon52} more than 
half a 
century ago. The basic simplicity of this model notwithstanding, it 
is actually quite appropriate for many realistic aspects of accretion 
processes, and certainly more than anything else, this model also allows 
a clear insight to be had into the related physics, much of which is 
often of quite an involved nature. As a result the spherically symmetric 
model is frequently the starting point from where it becomes possible 
to devise theoretical models of increasing physical complexity for 
accretion processes. 

While general questions related to astrophysical accretion have 
been addressed
and studied from various perspectives all along, it was not too long 
before Bondi's original treatment, carried out within the Newtonian
construct of space and time, was extended to embrace a general 
relativistic description of spherically symmetric accretion. An
early work in this regard was reported by Michel~\cite{mich72}, 
which was followed by a spate of later works, of which some took up 
various issues ranging from the stability of solutions to their exact
nature and critical 
aspects~\cite{bm76, beg78, bri80, mon80, mal99, ds01, mrd07}. 

Stability of spherically symmetric accretion has been studied long 
and studied variously~\cite{gar79, mon80, pso80, td92, ray03, gai06, rr07}. 
With especial regard to the stability of spherical accretion on to 
a black hole, Moncrief's study~\cite{mon80} has shown
no evidence of the development of any instability under the influence
of a linearized perturbation on the standing background flow. The 
treatment presented in this paper returns to the same general theme, 
but the methods applied here, and the motivation that has prompted 
them, are different. First of all, as opposed to Moncrief's approach 
of perturbing a scalar potential, whose gradient is prescribed to be 
the velocity of the ideal fluid, in this paper the perturbation 
scheme that has been adopted is centred around the continuity 
condition. This follows the procedure employed earlier by 
Petterson et al.~\cite{pso80} and Theuns \& David~\cite{td92}, 
respectively, in their studies
of spherical accretion in the non-relativistic regime. The stationary
solution of the continuity equation gives a first integral, which,
within a constant geometric factor, is actually the matter inflow
rate. The full description of the flow will imply that complete 
solutions of two coupled fields --- the velocity field and the 
density field --- will have to be obtained. Both these fields are
connected to the matter accretion rate, and, therefore, perturbing 
the accretion rate about its constant stationary value will lead  
to a wave equation for a single perturbed field that will also convey 
enough useful information on the stability of both the velocity and 
the density fields. 

While this is the general procedure that has been followed, the 
primary objective of the whole exercise has been to see whether 
or not through this perturbative technique it should be possible 
to establish an acoustic geometry for a flow that is at the 
same time described under fully general relativistic conditions. 
In recent years fluid analogue gravity has been a subject that has, 
extending over diverse types of fluid flows, called much attention
upon itself~\cite{un81, jacob91, un95, vis98, bil99, su02, das04,  
blv05, sbr05, us05, vol05, crd06, dbd06, vol06, rb07, rr07, rbh07}. 
In fluid 
dynamical processes a critical point is the point where the speed
of the bulk motion matches the speed of information propagation 
through the fluid. Depending on the direction of the flow, 
information of any event occurring in either the super-critical 
region of the fluid (where the bulk flow is faster than the speed 
with which any information can travel) or its sub-critical region 
(where information propagation overrides the bulk flow) will
not percolate into the other region, through a surface determined
by the critical condition. For an ideal fluid, therefore, the 
critical condition defines a barrier, which can 
be viewed as the fluid analogue of the event horizon of either
a black hole or a white hole, according to the direction in 
which the flow proceeds. It can be easily appreciated that for 
spherically symmetric accretion solutions the sonic surface 
defines the event horizon of an acoustic black hole. This 
feature has been studied extensively and understood well 
by now in both the Newtonian framework~\cite{crd06, rb07, rr07}  
as well as in the general relativistic framework~\cite{das04, dbd06}. 
The latter context is very interesting 
because it combines the metric properties of curved space-time
with similar properties of the fluid itself flowing in the 
same space-time.

This work purports to investigate the same analogue geometric  
properties of the flow, described in the Schwarzschild metric. 
A study along these very lines has been reported earlier by 
Bili\'c~\cite{bil99}, whose work, however, is based on the  
usual practice of studying a perturbation on a scalar potential 
function. In contrast, as it must be emphasized once again, 
the present work approaches the whole question from the 
viewpoint of the continuity condition in the flow. This difference 
of approach turns out to be significant because the result deriving
from the latter line of attack is negative, quite unlike what has 
been shown through the method used by Bili\'c~\cite{bil99}. 
This is rather curious 
because in the non-relativistic framework both paths can be shown
to lead to the same end~\cite{crd06, rb07, rr07}. 
It has been seen here that the coupling of the flow 
with the spherically symmetric geometry of space-time acts like a 
dissipative effect, and breaks down the Lorentz invariance that
should be necessary to define an analogue metric. The invariance
is restored in the Newtonian limit. By way of comparison and by
following the same mathematical prescription, one could invoke 
a similar feature that arises because of viscosity in shallow-layer 
incompressible fluid flows where information propagates as gravity 
waves~\cite{su02}. When viscosity is made to vanish, it becomes possible 
to establish an analogue black hole (or white hole) model for the flow
with the equivalent event horizon being set down by the condition
of the bulk flow speed matching the speed of gravity 
waves~\cite{sbr05, vol05, vol06, rbh07}. And so 
it is that when the conserved spherically symmetric accretion flow 
is decoupled from the space-time geometry in the Newtonian limit,
one could easily define a metric for an acoustic event horizon,
which would actually coincide with the sonic horizon of the flow. 

This whole aspect of the flow is manifested in its stability as
well. It has been seen that the stationary inflow solutions 
defined in the Schwarzschild metric are more stable under the 
effects of a linearized perturbation than what they should be 
in the Newtonian construct. Analogously mapped onto the properties
of the Newtonian limit, this is like the effect of viscous 
dissipation lending greater stability to the flow than what it
would have been for a perfect fluid. 

Finally, as an interesting aside, it has been shown that the steady
solution of the relativistic flow leads to the derivation of two
standard pseudo-Newtonian potentials which are often applied to 
mimic general relativistic effects in the Newtonain framework 
of space and time. 

\section{General relativistic equations for spherically 
symmetric accretion}
\label{sec2}

A complete description of a conserved spherically symmetric flow in 
Schwarzschild geometry will require some indispensible mathematical
relations. The first of these is the equation for the spherically
symmetric line element, which, in units of $c=1$, can be set down as 
\begin{equation}
\label{line}
{\mrm d} {\mathcal S}^2 = -f {\mrm d} t^2 + f^{-1} {\mrm d} r^2 
+ r^2 {\mrm d}\Omega^2
\end{equation} 
with $f\equiv f(r,t)$. This is to be followed by a relation for 
the momentum-energy tensor of a perfect fluid, given by~\cite{mis64} 
\begin{equation}
\label{stressenergy}
T^{\mu\nu} =\left(\epsilon + p\right)v^{\mu}v^{\nu}+pg^{\mu \nu}  
\end{equation} 
in which $p$ is the pressure, $\epsilon$ is the energy density 
and $v^{\mu}$ is the fluid four-velocity, which obeys the relation 
$v^{\mu}v_{\mu} = -1$. In tensorial notation the continuity 
condition can likewise be expressed as~\cite{shap83} 
\begin{equation}
\label{cont}
\left(\rho v^{\mu}\right)_{;\mu}=0
\end{equation} 
with $\rho$ being the particle number density of the perfect fluid. 
Some algebra following the use of the two conditions given by 
Eqs.~(\ref{line}) and~(\ref{cont}) will ultimately lead to a 
modified and explicit form of the continuity equation as 
\begin{equation} 
\label{modcon}
\frac{\prt}{\prt t}\left(\frac{\rho \sqrt{f + v^2}}{f} \right) 
+ \frac{1}{r^2}\frac{\prt}{\prt r}
\left(\rho v r^2\right) = 0 . 
\end{equation} 
This gives one relation connecting the radial flow velocity, $v$, and 
the local density, $\rho$, to each other. To solve for each of these
two fields explicitly, it should be necessary to obtain another
such relation. This can be derived from the momentum balance condition. 
To do so it should be necessary first to apply the requirement of 
energy-momentum conservation, ${T^{\mu \nu}}_{; \nu} = 0$, 
on Eq.~(\ref{stressenergy}). This will give
%\begin{widetext}
\begin{equation}
\label{enermomcon}
\left(\epsilon + p \right) \left({v^{\mu}}_{;\nu}v^{\mu} + 
v^{\nu}{v^{\nu}}_{;\nu} \right) + 
\left(\epsilon + p \right)_{,\nu}v^{\mu}v^{\nu} + 
g^{\mu \nu} p_{, \nu} = 0 
\end{equation}
%\end{widetext} 
with $\epsilon$ having to be defined separately through the 
thermodynamic relation~\cite{shap83}, 
\begin{equation}
\label{thermo}
\frac{{\mrm d}\epsilon}{{\mrm d}\rho} = \frac{\epsilon + p}
{\rho} + \rho T \frac{{\mrm d} s}{{\mrm d}\rho}
\end{equation}
in which $T$ is the temperature and $s$ is the specific entropy. 

A further definition that is necessary is that of the speed of 
sound, $a$, which, under isentropic conditions, is expressed 
in terms of the thermodynamic quantities, $\epsilon$ and $p$, 
as~\cite{shap83} 
\begin{equation}
\label{sound}
a^2 = \frac{\prt p}{\prt \epsilon} \bigg{\vert}_s . 
\end{equation} 
Making use of the foregoing definition, along with the condition 
of constant entropy in Eq.~(\ref{thermo}), achieved by setting 
${\mrm d} s =0$,  
and invoking the spherically symmetric line element from 
Eq.~(\ref{line}) once again, it should be a straightforward 
algebraic exercise to recast Eq.~(\ref{enermomcon}) in the form 
\begin{widetext}
\begin{equation}
\label{modenermom}
\frac{\sqrt{f + v^2}}{f} \frac{\prt v}{\prt t} +
v \frac{\prt v}{\prt r} + \frac{1}{2} \frac{\prt f}{\prt r}
- \frac{v \sqrt{f + v^2}}{f^2} \frac{\prt f}{\prt t} 
+ \frac{a^2}{\rho} \left[\frac{v \sqrt{f + v^2}}{f} 
\frac{\prt \rho}{\prt t}
+ \left(f + v^2\right) \frac{\prt \rho}{\prt r}\right] =0  
\end{equation}
\end{widetext} 
which, incidentally, bears a close resemblance with a similar 
equation derived by Misner \& Sharp~\cite{mis64} in their study 
of gravitational collapse. 

The pressure, $p$, is connected to the density, $\rho$, through 
a polytropic equation of state, $p = k\rho^{\gamma}$, with $k$
and $\gamma$ being constants, the latter being the polytropic 
exponent. With the help of this equation of state, it becomes 
easy to show from Eqs.~(\ref{thermo}) and~(\ref{sound}) that 
when the fluid is isentropic, there is a relation between $a$ 
and $\rho$ that can be written as  
\begin{equation}
\label{conden}
\rho =\left[\frac{a^2}{\gamma k
\left(1-na^2\right)}\right]^n 
\end{equation}
in which $n=(\gamma -1)^{-1}$, going by the usual definition 
of the polytropic index~\cite{chan39}. 
This connection between $a$ and $\rho$
now makes it possible to see Eq.~(\ref{modenermom}) as the 
second relation, after Eq.~(\ref{modcon}), that can be written 
entirely and explicitly in terms of $v$ and $\rho$. And so with the 
help of Eqs.~(\ref{modcon}),~(\ref{modenermom}) and~(\ref{conden}),
it should now be possible to establish a complete quantitative
description of the spherically symmetric flow defined within
the Schwarzschild metric. 

The stationary solutions of Eqs.~(\ref{modcon}) and~(\ref{modenermom})
are written as 
\begin{equation}
\label{intecon}
4 \pi {\bar{\mu}} \rho v r^2 = {\dot{m}}
\end{equation}
and 
\begin{equation}
\label{statmom}
\frac{1}{f + v^2}\frac{\mrm{d}}{{\mrm d}r}\left(f+v^2\right) = 
- \frac{2 a^2}{\rho}\frac{{\mrm d}\rho}{{\mrm d}r} , 
\end{equation}
respectively~\cite{shap83}. 
In the former solution, the integration constant, 
$\dot{m}$, is physically the matter flow rate, and $\bar{\mu}$ is the 
average density of particles in the flowing gas. While this solution 
is a direct first integral of the stationary continuity equation, 
the latter solution is not an integral solution. Nevertheless, it 
has certain interesting consequences which become very apparent in 
the non-relativistic limit. In this limit, both $v^2 \ll1$ and 
$a^2 \ll 1$, while $r \gg 2GM$. 

Now the static spherically symmetric metric
element will be given by $f = 1 - 2GM/r$, from Eq.~(\ref{line}). 
Therefore, in the 
non-relativistic limit the leading order solution that can be 
obtained from Eq.~(\ref{statmom}) is  
\begin{equation}
\label{stateuler}
\frac{1}{2}\frac{\mrm{d}}{{\mrm d}r}\left(v^2\right)
+ \phi^{\prm}(r) + 
\frac{a^2}{\rho}\frac{{\mrm d}\rho}{{\mrm d}r} = 0 
\end{equation} 
in which $\phi^{\prm} = f^{\prm}/2f$. 
What Eq.~(\ref{stateuler}) gives is the 
stationary Euler equation in the non-relativistic
limit, with $\phi$ being an effective potential driving the 
stationary flow. At large distances $\phi$ behaves like the 
classical Newtonian potential, but on length scales comparable
to $2GM$, there will be a deviation from the Newtonian behaviour. 

Frequently in astrophysics it becomes convenient to dispense 
with the full mathematical rigour of general relativity and 
instead, in a Newtonian framework, employ a prescription in which 
general relativistic effects could be represented by an effective 
potential. In this ``pseudo-Newtonian" approach, many such 
effective potentials have been suggested according to various 
specific requirements, of which two have been proposed 
by Artemova et al.~\cite{art96, ds01}. Interestingly enough, 
the functional behaviour 
of these two potentials could be derived from the form of $\phi$
implied in Eq.~(\ref{stateuler}). Written together, they are  
\begin{equation}
\label{pseupot}
\phi = \frac{1}{2} \ln\left(1 - \frac{2GM}{r}\right) \simeq
-1 + \sqrt{1 - \frac{2GM}{r}}  
\end{equation}
with the latter potential being seen to be actually a special case 
of the former. Both, of course, converge to the Newtonian limit on 
large length scales. 

While all these results can be derived from the stationary 
background flow, to have any understanding of their stability
under the effect of a linearized time-dependent perturbation, 
it will be necessary to go back to the two dynamic equations 
of the flow given by Eqs.~(\ref{modcon}) and~(\ref{modenermom}). 

\section{Linearized perturbations on stationary solutions}
\label{sec3}

For the purpose of carrying out the stability analysis of 
stationary solutions, a standard assumption that is being 
imposed is that the metric is static, i.e. $\prt f/\prt t =0$. 
This assumption is nothing unusual as far as accretion processes
are concerned, in which the flow is driven by the gravitational 
field of an external accretor. This will imply that the
gravitational field will be unchanging in time~\cite{bon52, mich72}. 

The perturbation scheme itself will be set down as 
$v(r,t) = v_0(r) + v^{\prm}(r,t)$ and 
$\rho(r,t) = \rho_0(r) + \rho^{\prm}(r,t)$ with the subscript
``0" indicating stationary values of $v$ and $\rho$, and the
primes indicating small time-dependent perturbations about
the stationary values. At this point, following the method
of Petterson et al.~\cite{pso80} and Theuns \& David~\cite{td92}, 
it will be expedient 
for the perturbative analysis to define a new variable, 
$\psi = \rho v r^2$, which, as it is very obvious from
Eq.~(\ref{modcon}), is closely associated with the matter flow 
rate, and whose stationary value, $\psi_0$, as it can be seen 
from Eq.~(\ref{intecon}), is a constant of the motion. The 
first-order fluctuations about this constant stationary value
can be expressed as 
\begin{equation}
\label{psifluc1}
\psi^{\prm} = \left(v_0 \rho^{\prm} + \rho_0 v^{\prm}\right)r^2 . 
\end{equation} 
Another such relation connecting $v^{\prm}$, $\rho^{\prm}$ and
$\psi^{\prm}$ can be derived from Eq.~(\ref{modcon}), and it will 
read as 
\begin{equation}
\label{psifluc2}
\frac{\sqrt{f + v_0^2}}{f} \frac{\prt \rho^{\prm}}{\prt t} 
+ \frac{\rho_0 v_0}{f\sqrt{f + v_0^2}} \frac{\prt v^{\prm}}{\prt t}
= - \frac{1}{r^2} \frac{\prt \psi^{\prm}}{\prt r} . 
\end{equation} 
With the help of the two foregoing equations it shall now be 
possible to express both $\rho^{\prm}$ and $v^{\prm}$ solely in 
terms of $\psi^{\prm}$. These will be given as 
\begin{equation}
\label{rhofluc}
\frac{\prt \rho^{\prm}}{\prt t} = - \frac{1}{r^2} 
\left(\frac{v_0}{f} \frac{\prt \psi^{\prm}}{\prt t} 
+ \sqrt{f + v_0^2} \frac{\prt \psi^{\prm}}{\prt r} \right) 
\end{equation}
and 
\begin{equation}
\label{vfluc}
\frac{\prt v^{\prm}}{\prt t} = \frac{\sqrt{f + v_0^2}}{\rho_0 r^2}
\left(\frac{\sqrt{f + v_0^2}}{f} \frac{\prt \psi^{\prm}}{\prt t} 
+ v_0 \frac{\prt \psi^{\prm}}{\prt r} \right) , 
\end{equation} 
respectively. 

The speed of sound, $a$, is connected to $\rho$ through Eq.~(\ref{conden}),
and so the perturbation in $\rho$ has to affect $a$ as well. This is to 
be written as $a^2 = a_0^2 + ({\mrm d}a_0^2/{\mrm d}\rho_0)\rho^{\prm}$.  
Once this has been done, the linearized first-order fluctuations about
the stationary momentum balance condition can be extracted from  
Eq.~(\ref{modenermom}) and written as 
\begin{widetext} 
\begin{equation}
\label{momfluc}
\frac{\sqrt{f + v_0^2}}{f} \frac{\prt v^{\prm}}{\prt t} + 
\frac{\prt}{\prt r}\left(v_0 v^{\prm} \right) + 
\frac{v_0 \sqrt{f + v_0^2}}{f} \frac{a_0^2}{\rho_0} 
\frac{\prt \rho^{\prm}}{\prt t} +
2v_0 \frac{\prt \rho_0}{\prt r} \frac{a_0^2}{\rho_0} v^{\prm} + 
\left(f + v_0^2 \right) \frac{\prt}{\prt r} 
\left(\frac{a_0^2}{\rho_0} \rho^{\prime} \right) = 0 . 
\end{equation}
\end{widetext} 
Partially differentiating Eq.~(\ref{momfluc}) with respect to time, 
and making use of 
Eqs.~(\ref{rhofluc}) and~(\ref{vfluc}) to eliminate $\rho^{\prm}$ and
$v^{\prm}$, respectively, will ultimately deliver a linearized equation
of motion for $\psi^{\prm}$ as 
\begin{widetext}
\begin{equation} 
\label{psieqnmot}
\frac{\prt}{\prt t} \left(\rho_0 h^{tt}
\frac{\prt \psi^{\prm}}{\prt t}
+ \rho_0 h^{tr}\frac{\prt \psi^{\prm}}{\prt r} \right) +  
\frac{\prt}{\prt r} 
\left(\rho_0 h^{rt}\frac{\prt \psi^{\prm}}{\prt t}
+ \rho_0 h^{rr}\frac{\prt \psi^{\prm}}{\prt r} \right) = 
(1 - 2a_0^2) \frac{{\mrm d}\rho_0}{{\mrm d}r} 
\left(h^{rt}\frac{\prt \psi^{\prm}}{\prt t}
+ h^{rr}\frac{\prt \psi^{\prm}}{\prt r} \right)
\end{equation}
\end{widetext} 
with the coefficients, $h^{\alpha \beta}$, having to be read from 
\begin{displaymath}
\label{htt}
h^{tt} = \frac{v_0 \sqrt{f + v_0^2}}{f^2}
\left(f + v_0^2 - v_0^2 a_0^2 \right) , 
\end{displaymath} 
\begin{displaymath}
\label{htr}
h^{tr} = h^{rt} = \frac{v_0^2 \left(f + v_0^2 \right)}{f}
(1 - a_0^2)
\end{displaymath} 
and 
\begin{displaymath}
\label{hrr}
h^{rr} = v_0 \sqrt{f + v_0^2} \left[v_0^2 - \left(f + v_0^2 \right)
a_0^2 \right] .  
\end{displaymath} 

At this stage it would be very much instructive to examine the 
features of Eq.~(\ref{psieqnmot}) in the non-relativistic limit,
where both $v_0^2$ and $a_0^2$ are vanishingly small compared to 
unity while $f$ itself assumes the value of unity. As a 
consequence, in the non-relativistic limit it will become eminently
possible to reduce Eq.~(\ref{psieqnmot}) to a compact form given by 
\begin{equation}
\label{compact} 
{\prt}_{\alpha}\left(h^{\alpha \beta} {\prt}_{\beta} 
\psi^{\prm} \right) = 0
\end{equation} 
with the Greek indices $\alpha$ and $\beta$ running over $t$ and $r$, 
as they did for the fully relativistic case earlier. 

In Lorentzian geometry the d'Alembertian for a scalar field in curved 
space is given in terms of the metric $g_{\mu \nu}$ by~\cite{vis98} 
\begin{equation}
\label{alem} 
\Delta \varphi\equiv \frac{1}{\sqrt{-g}}
\partial_\mu\left({\sqrt{-g}}\, g^{\mu\nu}\partial_\nu\varphi\right)
\end{equation}
with $g^{\mu\nu}$ being the inverse of the matrix implied
by $g_{\mu\nu}$. Comparing Eq.~(\ref{compact})
with Eq.~(\ref{alem}),
it would be tempting to look for an exact equivalence between
$h^{\alpha \beta}$ (in the non-relativistic limit) and 
$\sqrt{-g}\, g^{\mu\nu}$.
This, however, cannot be done in a general sense. What can be
appreciated, nevertheless, is that Eq.~(\ref{compact}) gives
a relation for $\psi^{\prm}$ which is of the type given by
Eq.~(\ref{alem}). The metrical part of Eq.~(\ref{compact}),
as given by the values of $h^{\alpha \beta}$, may then be
extracted, and its inverse will incorporate the notion of a 
sonic horizon of an acoustic black hole when $v_0^2 = a_0^2$.
This point of view has some features similar to the metric 
of a wave equation for a scalar field in curved space-time, 
obtained through a somewhat different approach, in which the 
velocity of an irrotational, inviscid and barotropic fluid 
flow is first represented as the gradient of a scalar function, 
and then a perturbation is imposed on this scalar 
function~\cite{vis98, blv05}.

While all this similarity is undoubtedly pleasing to note in
the non-relativistic limit, it is also quite obvious from 
Eq.~(\ref{psieqnmot}) that in the fully general relativistic
treatment there is a breakdown of the symmetry that leads to 
the devising of an analogue gravity model for the fluid flow. 
This is at striking variance with the conclusions arrived at 
by Bili\'c~\cite{bil99}. It is not very difficult to discern that 
the difference arises because of the way the perturbative 
studies have been prescribed in the two cases --- the scalar
potential approach of Bili\'c~\cite{bil99}, and the continuity 
equation approach in the present case. 

Having said so, it must nevertheless be pointed out that in some 
respects at least the fluid analogue approach is not entirely 
lost even for the general relativistic flow being studied
here. Speaking by analogy, in the shallow layer flow of a 
perfect liquid (without any viscosity) it has been shown 
that an analogue black hole or white hole model is very much 
a fact~\cite{su02}. Viscous dissipation in the 
flow, on the other hand, adversely affects the invariance
that makes the analogue model possible, even as it helps   
the hydraulic jump phenomenon itself to happen~\cite{sbr05, rbh07}. 
And in a like fashion  
while it is easy to recognize the concept of fluid analogue gravity 
for spherically symmetric accretion in the non-relativistic 
situation, the coupling of the geometry of space-time with 
the perturbed field in the fully general relativistic scenario
acts in the manner of a dissipative effect. At least through 
the continuity equation approach this precludes any 
hope of building an acoustic black hole model within the general 
relativistic flow, but at the same time this has a more 
favourable role to play as regards the stability of the 
stationary solutions.  

\section{Stability analysis : Standing waves} 
\label{sec4} 

In treating the perturbation as a standing wave it will be 
essential to identify proper boundary conditions to constrain 
the wave at two spatial points. 
Between these two points the 
properties of the standing wave could be studied. 
One boundary
condition for the perturbation is conveniently fixed at the 
outer boundary of the stationary flow itself, where any solution
naturally decays out. So should any perturbation imposed on it. 

In determining an appropriate inner boundary condition, on the 
other hand, one encounters much greater difficulties. The nature
of the accretor itself has an influential role to play in this
matter. If it is a black hole then the infalling matter has to
cross the event horizon at the maximum possible rate~\cite{nov73}, 
and the only feasible solution should be a transonic one, which 
will smoothly pass through a singular point in the flow~\cite{skc90}. 
On the other hand, if the accretor is a compact object with a
physical surface, then the inner boundary condition becomes 
subject to many complications, which in turn will leave its
imprint on the character of an inflow solution in the vicinity
of the stellar boundary. The solution could either be transonic
or shocked (discontinuous) subsonic or continuously subsonic. 
These distinct aspects have been discussed at length by 
Petterson et al.~\cite{pso80} and Theuns \& David~\cite{td92}. 
In this context Moncrief~\cite{mon80} 
has pointed out that as long as the stellar surface is enclosed
within the sonic surface, his perturbative arguments hold both
for a black hole as well as a compact star. 

The standing wave analysis that has been pursued here will, of
necessity, require the background solution to be continuous 
everywhere and globally well behaved. Besides this the wave
will also have to die out at the two chosen boundaries. Now 
the only solutions that will meet all these requirements in 
a general sense are the purely subsonic solutions. While these
may not entirely be representative of the precise manner of 
infall in the general relativistic scenario, a mathematical 
study of their stability will reveal the true extent of the 
influence that the coupling of the flow with the geometry 
of space-time will have vis-a-vis what it is in the Newtonian
flat space-time limit. Stability of subsonic flows in the latter 
situation has been studied thoroughly by Petterson et al.~\cite{pso80}  
and Theuns \& David~\cite{td92} .  

The solution  
$\psi^{\prm}(r,t)=p(r)\exp\left(-{\mrm i}\omega t\right)$ 
is to be applied in Eq.~(\ref{psieqnmot}) and the resulting 
expression multiplied throughout by $p$. This will lead to 
\begin{widetext}
\begin{equation}
\label{interpert1}
h^{tt} p^2 \omega^2 + {\mrm i} \left\{ \frac{\mrm d}{{\mrm d} r} 
\left(h^{tr} p^2 \right) - h^{rt} p^2 
\frac{\mrm d}{{\mrm d}r}
\left[\ln \left(f + v_0^2 \right) \right]\right\}\omega
+ \frac{h^{rr}}{f + v_0^2} \frac{{\mrm d}p}{{\mrm d}r} 
\frac{\mrm d}{{\mrm d}r}\left[p \left(f + v_0^2 \right)\right] 
- \frac{\mrm d}{{\mrm d} r}\left(h^{rr}p
\frac{{\mrm d} p}{{\mrm d} r}\right)
= 0 
\end{equation}
\end{widetext} 
which will then have to be integrated over the entire spatial
range within which the standing wave is continuously distributed. 
At the two boundaries of this range the amplitude of the wave
is required to vanish. So the ``surface" terms obtained from 
integrating Eq.(\ref{interpert1}) will also have to vanish at the
boundaries. This will leave a residual quadratic equation which 
will be a dispersion relation for $\omega$. This relation will 
be in the form 
\begin{equation}
\label{disper}
A \omega^2 - 2 {\mrm i} B \omega + C = 0
\end{equation}
with the three coefficients above being read as 
\begin{displaymath}
\label{aae}
A = \int h^{tt} p^2 {\mrm d} r , 
\end{displaymath} 
\begin{displaymath}
\label{bee}
B = \int \frac{h^{rt}p^2}{2} \frac{\mrm d}{{\mrm d}r}
\left[ \ln \left(f + v_0^2 \right) \right] {\mrm d} r
\end{displaymath} 
and 
\begin{displaymath}
\label{cee} 
C = \int \frac{h^{rr}}{f + v_0^2} \frac{{\mrm d}p}{{\mrm d}r}
\frac{\mrm d}{{\mrm d}r}\left[p \left(f + v_0^2 \right) \right]
{\mrm d} r , 
\end{displaymath} 
respectively. 

Under the condition that for inflow solutions $v<0$ and 
$({\mrm d}\rho_0/{\mrm d}r) <0$, it shall be easy to verify 
that $(B/A) <0$ by referring to Eq.~(\ref{statmom}). 
Therefore, the discriminant of the solution 
of Eq.~(\ref{disper}) will hold the key regarding the stability
of the standing wave. Once again, for subsonic solutions, it 
can be argued that $(C/A)<0$. And so, 
if $\vert C/A \vert > (B/A)^2$, it will imply that the 
time-dependent part of the standing wave, given by 
$\exp({\mrm i}\omega t)$, will have an oscillatory nature
with the amplitude of the oscillation being damped in time. 
Hence the background solution will be stable. If, on the 
other hand, $\vert C/A \vert < (B/A)^2$, then there will 
be two real roots of $\omega$, both negative, indicating that 
the perturbation will be overdamped. So, one way or the 
other, the amplitude of the standing wave will be damped, 
lending stability to the stationary background solution 
in strong measure. 

This is a rather intriguing state of affairs indeed. The 
decay of the amplitude of the standing wave would imply that 
something in the nature of a dissipative effect is active
in what is otherwise a conservative system. For a conserved
flow in the Newtonian regime, Petterson et al.~\cite{pso80} 
have shown 
that the perturbation will have a constant amplitude. Any 
decay of the amplitude of the standing wave could only be 
reproduced when one accounts for viscosity in the flow~\cite{ray03}.
Coming back to the conserved general relativistic case, 
the only possible explanation for the decaying behaviour 
can be that the coupling of the flow with the geometry of 
space-time acts in the manner of an ``effective" dissipation. 
And consistent with this 
line of thinking it can also be shown that in the Newtonian 
limit one does indeed regain the expected constancy of the 
amplitude of the standing waves. 

\section{Stability analysis : Travelling waves} 
\label{sec5} 

The manner in which the stability of the flow is influenced by 
its coupling with the space-time metric could also be examined by 
fashioning the perturbation to be a high-frequency travelling wave. 
A comparison could then be made with the corresponding analysis
carried out in the Newtonian structure of space and time by Petterson 
et al.~\cite{pso80} who argued that the travelling waves could cause 
a growth
in the fluctuations on the flow rate, but would not drive the 
background flow from it stationary profile. To this extent 
the stability of the flow should be preserved.    

High-frequency travelling waves are to be first defined precisely
in the present context by the fact that their wavelength should
be much smaller than the Schwarzschild radius of the black hole.
This will imply that the frequency, $\omega$, should be 
correspondingly large. With this restriction on $\omega$, the
spatial part of the perturbation can then be prescribed in terms 
of a power series as
\begin{equation}
\label{power}
p_{\omega}(r) = \exp \left[ \sum_{l=-1}^\infty \omega^{-l}
k_{l}(r) \right]
\end{equation}
and this is then to be applied to a slightly modified rendering 
of Eq.~(\ref{interpert1}) which goes as 
\begin{widetext}
\begin{equation}
\label{interpert2}
h^{rr}\frac{{\mrm d}^2 p}{{\mrm d}r^2} + 
\left\{\frac{{\mrm d}h^{rr}}{{\mrm d}r} - 2 {\mrm i} \omega h^{tr}
- h^{rr}\frac{\mrm d}{{\mrm d}r}\left[\ln\left(f + v_0^2 \right)\right]
\right\} \frac{{\mrm d} p}{{\mrm d} r} - \left\{ \omega^2 h^{tt}
+ {\mrm i} \omega \frac{{\mrm d}h^{rt}}{{\mrm d}r} - {\mrm i} \omega
h^{rt} \frac{\mrm d}{{\mrm d}r}\left[\ln\left(f + v_0^2 \right)\right]
\right\}p = 0 . 
\end{equation}
\end{widetext}

From the result of this extended algebraic exercise all coefficients
of $\omega^2$ are to be collected and their sum is to be set to zero. 
This will give a solution for $k_{-1}$, which will look like
\begin{equation}
\label{kayminus1}
k_{-1} = {\mrm i} \int \left(h^{rr}\right)^{-1} 
\left[h^{tr}\pm\sqrt{\left(h^{tr}\right)^2-h^{rr}h^{tt}}\right] \, 
{\mrm d}r . 
\end{equation} 
Similarly summing up all the coefficients of $\omega$ to be zero, 
and applying the value of $k_{-1}$, as Eq.~(\ref{kayminus1}) gives it, 
will deliver a solution for $k_0$ as 
\begin{equation}
\label{kaynot}
k_0 = \ln \left\{\left(f + v_0^2\right)\left[
\sqrt{\left(h^{tr}\right)^2-h^{rr}h^{tt}}\right]^{-1}\right\}^{1/2} . 
\end{equation} 
Likewise, the solution of $k_1$ could be found by setting
the sum of the coefficients of $\omega^0$ to be zero. In terms of 
$k_{-1}$ and $k_0$ this can be expressed as
\begin{widetext}
\begin{equation}
\label{kay1}
2 \left(h^{rr} \frac{{\mrm d}k_{-1}}{{\mrm d}r} - {\mrm i} h^{tr}\right)
\frac{{\mrm d}k_1}{{\mrm d}r} 
+ \frac{\mrm d}{{\mrm d}r} \left(h^{rr}\frac{{\mrm d}k_0}{{\mrm d}r}\right) 
+ h^{rr}\frac{{\mrm d}k_0}{{\mrm d}r} \frac{\mrm d}{{\mrm d}r} \left[k_0
- \ln\left(f + v_0^2 \right) \right] = 0 . 
\end{equation}
\end{widetext} 

For reasons of self-consistency it shall be necessary at this stage to 
show that successive terms in the power series given by Eq.~(\ref{power})
will obey the requirement that 
$\omega^{-l}\vert k_l(r) \vert \gg \omega^{-(l+1)}\vert k_{l+1}(r)\vert$,
which will also imply that the power series will converge 
rapidly with increasing $l$, and so it can be truncated 
after the first few terms. Mindful of the fact that $\omega$ is large,
the first three terms, involving $k_{-1}$, $k_0$ and $k_1$, respectively,
conform to this self-consistency requirement. A simple asymptotic check
suffices to show $k_{-1} \sim r$, $k_0 \sim \ln r$ and $k_1 \sim r^{-1}$.  
In any case both $k_{-1}$ and $k_1$ make contributions to the phase of
the travelling wave. In this linearized treatment, therefore, the most
conspicuous contribution to the amplitude comes from $k_0$, and any 
impression of the stability of the flow can be unambiguously derived 
from this term only. 

Using only the solutions of $k_{-1}$ and $k_0$, the dominant properties
of the perturbation could be set down as  
\begin{widetext}
\begin{equation}
\label{psisol}
\psi^{\prm}(r,t) \simeq \xi_{\pm} \left[\frac{f + v_0^2}
{\sqrt{\left(h^{tr}\right)^2-h^{rr}h^{tt}}} \right]^{1/2}
\exp\left\{{\mrm i} \omega \int \left(h^{rr}\right)^{-1}
\left[h^{tr}\pm\sqrt{\left(h^{tr}\right)^2-h^{rr}h^{tt}}\right] \,
{\mrm d}r \right\} e^{{\mrm i} \omega t} 
\end{equation} 
\end{widetext} 
with $\xi$ being a constant, and with the positive and negative signs, 
placed together, indicating a superposition of incoming (corresponding 
to the negative sign) and outgoing (corresponding to the positive sign) 
travelling waves, respectively. 

It will now be very much worthwhile 
to scrutinize the expression for $k_0$ more closely and see how the 
geometry of space-time makes its contribution to stability. Using 
the derived values of $h^{tt}$, $h^{tr}$ and $h^{rr}$ it can be shown
that $k_0$ picks up a term that goes as $\ln f$. By virtue of the fact 
that $f <1$, the logarithm of $f$ will be negative, and consequently 
its effect on $k_0$ would be to detract from its value in the Newtonian 
limit (where $f=1$). And so where general relativistic
effects will have to be accounted for, the amplitude of the travelling
waves will become subdued. Once again it will not be difficult
to see that this effect is owed entirely to the coupling between the 
flow and the curvature of space-time. 

To delve into some more details, the amplitude of the perturbation 
could be recast very simply as 
\begin{equation}
\label{ampli}
\vert \psi^{\prm} \vert \sim 
\left(\frac{f + v_0^2}{a_0^2 v_0^2}\right)^{1/4}
\end{equation}
a form which is quite helpful in shedding a clear light on the 
asymptotic behaviour. In the Newtonian limit, it is obvious that
$\vert \psi^{\prm} \vert \sim (a_0 v_0)^{-1/2}$, a state of affairs
whose stability has been cogently argued for by Petterson et 
al.~\cite{pso80}. An equal measure of stability is to be seen 
near the event horizon as well, with a smooth passage for the 
travelling wave through the sonic region somewhere in between. 
After this it becomes easy to argue for the asymptotic
stability of the steady solutions. 

\section{Concluding remarks} 
\label{sec6}

Some parting comments would well be in order. It has been shown 
already that the manner in which the perturbative study has proceeded,
beginning with the continuity condition of the flow, has led to a 
failure in establishing an acoustic geometry for general relativistic
spherically symmetric accretion. On the other hand, as far as 
stability is concerned, this approach has been in perfect qualitative
conformity with earlier studies which have dwelt on the question of 
the stability of the flow solutions. Indeed as regards stability in
particular, it has been argued and shown that general relativistic
effects enhance the stability of the stationary solutions. This fact 
actually opens up an interesting possibility. 

Schwarzschild space-time defines a geometry of closed curvature. A
standing wave in this geometry exhibits a damping of its amplitude, 
and so a stable behaviour is implied. In the Newtonian limit, the 
standing waves continue to have a constant amplitude. In this respect
the behaviour may once again be viewed to be stable. However, the 
situation could become radically different in the geometry of open
curvature. One might conjecture that in this case, simply because 
of the nature of the geometry, there will be an unstable behaviour, 
manifested through a growth in the amplitude of a standing wave. 
Indeed, this speculation is not without its foundations. In a 
different context, but with good consonance, similar features 
are to be seen in CMB anisotropy in compact hyperbolic 
spaces~\cite{bps1, bps2}. 

\begin{acknowledgments}
This research has made use of NASA's Astrophysics Data System.
Tapan Naskar would like to thank CSIR, Government of India for 
a research fellowship. 
Nabajit Chakrabarty is grateful to DGM, IMD and Govt. of India for granting study leave (DGM order No. A-24036/I/05-E(2), dated 02.07.2007).
The authors acknowledge some useful 
discussions with D. Kothawala, G. Mahajan, S. Roychowdhury and 
T. Souradeep. Gratitude is also to be expressed to A. K. Kembhavi 
for his support in many respects. 
\end{acknowledgments}

\bibliography{ncbr_aug}

\begin{thebibliography}{39}
\expandafter\ifx\csname natexlab\endcsname\relax\def\natexlab#1{#1}\fi
\expandafter\ifx\csname bibnamefont\endcsname\relax
  \def\bibnamefont#1{#1}\fi
\expandafter\ifx\csname bibfnamefont\endcsname\relax
  \def\bibfnamefont#1{#1}\fi
\expandafter\ifx\csname citenamefont\endcsname\relax
  \def\citenamefont#1{#1}\fi
\expandafter\ifx\csname url\endcsname\relax
  \def\url#1{\texttt{#1}}\fi
\expandafter\ifx\csname urlprefix\endcsname\relax\def\urlprefix{URL }\fi
\providecommand{\bibinfo}[2]{#2}
\providecommand{\eprint}[2][]{\url{#2}}

\bibitem[{\citenamefont{Bondi}(1952)}]{bon52}
\bibinfo{author}{\bibfnamefont{H.}~\bibnamefont{Bondi}}, \bibinfo{journal}{Mon.
  Not. R. Astron. Soc.} \textbf{\bibinfo{volume}{112}}, \bibinfo{pages}{195}
  (\bibinfo{year}{1952}).

\bibitem[{\citenamefont{Michel}(1972)}]{mich72}
\bibinfo{author}{\bibfnamefont{F.~C.} \bibnamefont{Michel}},
  \bibinfo{journal}{Astrophys. Space Sci.} \textbf{\bibinfo{volume}{15}},
  \bibinfo{pages}{153} (\bibinfo{year}{1972}).

\bibitem[{\citenamefont{Blumenthal and Mathews}(1976)}]{bm76}
\bibinfo{author}{\bibfnamefont{G.~R.} \bibnamefont{Blumenthal}}
  \bibnamefont{and} \bibinfo{author}{\bibfnamefont{W.~G.}
  \bibnamefont{Mathews}}, \bibinfo{journal}{Astrophys. J.}
  \textbf{\bibinfo{volume}{203}}, \bibinfo{pages}{714} (\bibinfo{year}{1976}).

\bibitem[{\citenamefont{Begelman}(1978)}]{beg78}
\bibinfo{author}{\bibfnamefont{M.~C.} \bibnamefont{Begelman}},
  \bibinfo{journal}{Astron. Astrophys.} \textbf{\bibinfo{volume}{70}},
  \bibinfo{pages}{53} (\bibinfo{year}{1978}).

\bibitem[{\citenamefont{Brinkmann}(1980)}]{bri80}
\bibinfo{author}{\bibfnamefont{W.}~\bibnamefont{Brinkmann}},
  \bibinfo{journal}{Astron. Astrophys.} \textbf{\bibinfo{volume}{85}},
  \bibinfo{pages}{146} (\bibinfo{year}{1980}).

\bibitem[{\citenamefont{Moncrief}(1980)}]{mon80}
\bibinfo{author}{\bibfnamefont{V.}~\bibnamefont{Moncrief}},
  \bibinfo{journal}{Astrophys. J.} \textbf{\bibinfo{volume}{235}},
  \bibinfo{pages}{1038} (\bibinfo{year}{1980}).

\bibitem[{\citenamefont{Malec}(1999)}]{mal99}
\bibinfo{author}{\bibfnamefont{E.}~\bibnamefont{Malec}},
  \bibinfo{journal}{Phys. Rev. D} \textbf{\bibinfo{volume}{60}},
  \bibinfo{pages}{104043} (\bibinfo{year}{1999}).

\bibitem[{\citenamefont{Das and Sarkar}(2001)}]{ds01}
\bibinfo{author}{\bibfnamefont{T.~K.} \bibnamefont{Das}} \bibnamefont{and}
  \bibinfo{author}{\bibfnamefont{A.}~\bibnamefont{Sarkar}},
  \bibinfo{journal}{Astron. Astrophys.} \textbf{\bibinfo{volume}{374}},
  \bibinfo{pages}{1150} (\bibinfo{year}{2001}).

\bibitem[{\citenamefont{Mandal et~al.}(2007)\citenamefont{Mandal, Ray, and
  Das}}]{mrd07}
\bibinfo{author}{\bibfnamefont{I.}~\bibnamefont{Mandal}},
  \bibinfo{author}{\bibfnamefont{A.~K.} \bibnamefont{Ray}}, \bibnamefont{and}
  \bibinfo{author}{\bibfnamefont{T.~K.} \bibnamefont{Das}},
  \bibinfo{journal}{Mon. Not. R. Astron. Soc.} \textbf{\bibinfo{volume}{378}},
  \bibinfo{pages}{1400} (\bibinfo{year}{2007}).

\bibitem[{\citenamefont{Garlick}(1979)}]{gar79}
\bibinfo{author}{\bibfnamefont{A.~R.} \bibnamefont{Garlick}},
  \bibinfo{journal}{Astron. Astrophys.} \textbf{\bibinfo{volume}{73}},
  \bibinfo{pages}{171} (\bibinfo{year}{1979}).

\bibitem[{\citenamefont{Petterson et~al.}(1980)\citenamefont{Petterson, Silk,
  and Ostriker}}]{pso80}
\bibinfo{author}{\bibfnamefont{J.~A.} \bibnamefont{Petterson}},
  \bibinfo{author}{\bibfnamefont{J.}~\bibnamefont{Silk}}, \bibnamefont{and}
  \bibinfo{author}{\bibfnamefont{J.~P.} \bibnamefont{Ostriker}},
  \bibinfo{journal}{Mon. Not. R. Astron. Soc.} \textbf{\bibinfo{volume}{191}},
  \bibinfo{pages}{571} (\bibinfo{year}{1980}).

\bibitem[{\citenamefont{Theuns and David}(1992)}]{td92}
\bibinfo{author}{\bibfnamefont{T.}~\bibnamefont{Theuns}} \bibnamefont{and}
  \bibinfo{author}{\bibfnamefont{M.}~\bibnamefont{David}},
  \bibinfo{journal}{Astrophys. J.} \textbf{\bibinfo{volume}{384}},
  \bibinfo{pages}{587} (\bibinfo{year}{1992}).

\bibitem[{\citenamefont{Ray}(2003)}]{ray03}
\bibinfo{author}{\bibfnamefont{A.~K.} \bibnamefont{Ray}},
  \bibinfo{journal}{Mon. Not. R. Astron. Soc.} \textbf{\bibinfo{volume}{344}},
  \bibinfo{pages}{1085} (\bibinfo{year}{2003}).

\bibitem[{\citenamefont{Gaite}(2006)}]{gai06}
\bibinfo{author}{\bibfnamefont{J.}~\bibnamefont{Gaite}},
  \bibinfo{journal}{Astron. Astrophys.} \textbf{\bibinfo{volume}{449}},
  \bibinfo{pages}{861} (\bibinfo{year}{2006}).

\bibitem[{\citenamefont{Roy and Ray}(2007)}]{rr07}
\bibinfo{author}{\bibfnamefont{N.}~\bibnamefont{Roy}} \bibnamefont{and}
  \bibinfo{author}{\bibfnamefont{A.~K.} \bibnamefont{Ray}},
  \bibinfo{journal}{Mon. Not. R. Astron. Soc.} \textbf{\bibinfo{volume}{(To
  appear)}} (\bibinfo{year}{2007}).

\bibitem[{\citenamefont{Unruh}(1981)}]{un81}
\bibinfo{author}{\bibfnamefont{W.}~\bibnamefont{Unruh}},
  \bibinfo{journal}{Phys. Rev. Lett.} \textbf{\bibinfo{volume}{46}},
  \bibinfo{pages}{1351} (\bibinfo{year}{1981}).

\bibitem[{\citenamefont{Jacobson}(1991)}]{jacob91}
\bibinfo{author}{\bibfnamefont{T.}~\bibnamefont{Jacobson}},
  \bibinfo{journal}{Phys. Rev. D} \textbf{\bibinfo{volume}{44}},
  \bibinfo{pages}{1731} (\bibinfo{year}{1991}).

\bibitem[{\citenamefont{Unruh}(1995)}]{un95}
\bibinfo{author}{\bibfnamefont{W.}~\bibnamefont{Unruh}},
  \bibinfo{journal}{Phys. Rev. D} \textbf{\bibinfo{volume}{51}},
  \bibinfo{pages}{2827} (\bibinfo{year}{1995}).

\bibitem[{\citenamefont{Visser}(1998)}]{vis98}
\bibinfo{author}{\bibfnamefont{M.}~\bibnamefont{Visser}},
  \bibinfo{journal}{Class. Quantum Grav.} \textbf{\bibinfo{volume}{15}},
  \bibinfo{pages}{1767} (\bibinfo{year}{1998}).

\bibitem[{\citenamefont{Bili\'c}(1999)}]{bil99}
\bibinfo{author}{\bibfnamefont{N.}~\bibnamefont{Bili\'c}},
  \bibinfo{journal}{Class. Quantum Grav.} \textbf{\bibinfo{volume}{16}},
  \bibinfo{pages}{3953} (\bibinfo{year}{1999}).

\bibitem[{\citenamefont{Sch{\"u}tzhold and Unruh}(2002)}]{su02}
\bibinfo{author}{\bibfnamefont{R.}~\bibnamefont{Sch{\"u}tzhold}}
  \bibnamefont{and} \bibinfo{author}{\bibfnamefont{W.}~\bibnamefont{Unruh}},
  \bibinfo{journal}{Phys. Rev. D} \textbf{\bibinfo{volume}{66}},
  \bibinfo{pages}{044019} (\bibinfo{year}{2002}).

\bibitem[{\citenamefont{Das}(2004)}]{das04}
\bibinfo{author}{\bibfnamefont{T.~K.} \bibnamefont{Das}},
  \bibinfo{journal}{Class. Quantum Grav.} \textbf{\bibinfo{volume}{21}},
  \bibinfo{pages}{5253} (\bibinfo{year}{2004}).

\bibitem[{\citenamefont{Barcel{\'o} et~al.}()\citenamefont{Barcel{\'o},
  Liberati, and Visser}}]{blv05}
\bibinfo{author}{\bibfnamefont{C.}~\bibnamefont{Barcel{\'o}}},
  \bibinfo{author}{\bibfnamefont{S.}~\bibnamefont{Liberati}}, \bibnamefont{and}
  \bibinfo{author}{\bibfnamefont{M.}~\bibnamefont{Visser}},
  \eprint{gr-qc/0505065}.

\bibitem[{\citenamefont{Singha et~al.}(2005)\citenamefont{Singha,
  Bhattacharjee, and Ray}}]{sbr05}
\bibinfo{author}{\bibfnamefont{S.~B.} \bibnamefont{Singha}},
  \bibinfo{author}{\bibfnamefont{J.~K.} \bibnamefont{Bhattacharjee}},
  \bibnamefont{and} \bibinfo{author}{\bibfnamefont{A.~K.} \bibnamefont{Ray}},
  \bibinfo{journal}{Eur. Phys. J. B} \textbf{\bibinfo{volume}{48}},
  \bibinfo{pages}{417} (\bibinfo{year}{2005}).

\bibitem[{\citenamefont{Unruh and Sch{\"u}tzhold}(2005)}]{us05}
\bibinfo{author}{\bibfnamefont{W.}~\bibnamefont{Unruh}} \bibnamefont{and}
  \bibinfo{author}{\bibfnamefont{R.}~\bibnamefont{Sch{\"u}tzhold}},
  \bibinfo{journal}{Phys. Rev. D} \textbf{\bibinfo{volume}{71}},
  \bibinfo{pages}{024028} (\bibinfo{year}{2005}).

\bibitem[{\citenamefont{Volovik}(2005)}]{vol05}
\bibinfo{author}{\bibfnamefont{G.~E.} \bibnamefont{Volovik}},
  \bibinfo{journal}{JETP Letters} \textbf{\bibinfo{volume}{82}},
  \bibinfo{pages}{624} (\bibinfo{year}{2005}).

\bibitem[{\citenamefont{Chaudhury et~al.}(2006)\citenamefont{Chaudhury, Ray,
  and Das}}]{crd06}
\bibinfo{author}{\bibfnamefont{S.}~\bibnamefont{Chaudhury}},
  \bibinfo{author}{\bibfnamefont{A.~K.} \bibnamefont{Ray}}, \bibnamefont{and}
  \bibinfo{author}{\bibfnamefont{T.~K.} \bibnamefont{Das}},
  \bibinfo{journal}{Mon. Not. R. Astron. Soc.} \textbf{\bibinfo{volume}{373}},
  \bibinfo{pages}{146} (\bibinfo{year}{2006}).

\bibitem[{\citenamefont{Das et~al.}()\citenamefont{Das, Bili{\'c}, and
  Dasgupta}}]{dbd06}
\bibinfo{author}{\bibfnamefont{T.~K.} \bibnamefont{Das}},
  \bibinfo{author}{\bibfnamefont{N.}~\bibnamefont{Bili{\'c}}},
  \bibnamefont{and} \bibinfo{author}{\bibfnamefont{S.}~\bibnamefont{Dasgupta}},
  \eprint{astro-ph/0604477}.

\bibitem[{\citenamefont{Volovik}(2006)}]{vol06}
\bibinfo{author}{\bibfnamefont{G.~E.} \bibnamefont{Volovik}},
  \bibinfo{journal}{J. Low Temp. Phys.} \textbf{\bibinfo{volume}{145}},
  \bibinfo{pages}{337} (\bibinfo{year}{2006}).

\bibitem[{\citenamefont{Ray and Bhattacharjee}(2007{\natexlab{a}})}]{rb07}
\bibinfo{author}{\bibfnamefont{A.~K.} \bibnamefont{Ray}} \bibnamefont{and}
  \bibinfo{author}{\bibfnamefont{J.~K.} \bibnamefont{Bhattacharjee}},
  \bibinfo{journal}{Class. Quantum Grav.} \textbf{\bibinfo{volume}{24}},
  \bibinfo{pages}{1479} (\bibinfo{year}{2007}{\natexlab{a}}).

\bibitem[{\citenamefont{Ray and Bhattacharjee}(2007{\natexlab{b}})}]{rbh07}
\bibinfo{author}{\bibfnamefont{A.~K.} \bibnamefont{Ray}} \bibnamefont{and}
  \bibinfo{author}{\bibfnamefont{J.~K.} \bibnamefont{Bhattacharjee}},
  \bibinfo{journal}{Phys. Lett. A} \textbf{\bibinfo{volume}{(To appear)}}
  (\bibinfo{year}{2007}{\natexlab{b}}).

\bibitem[{\citenamefont{Misner and Sharp}(1964)}]{mis64}
\bibinfo{author}{\bibfnamefont{C.~W.} \bibnamefont{Misner}} \bibnamefont{and}
  \bibinfo{author}{\bibfnamefont{D.~H.} \bibnamefont{Sharp}},
  \bibinfo{journal}{Phys. Rev.} \textbf{\bibinfo{volume}{136B}},
  \bibinfo{pages}{571} (\bibinfo{year}{1964}).

\bibitem[{\citenamefont{Shapiro and Teukolsky}(1983)}]{shap83}
\bibinfo{author}{\bibfnamefont{S.~L.} \bibnamefont{Shapiro}} \bibnamefont{and}
  \bibinfo{author}{\bibfnamefont{S.~A.} \bibnamefont{Teukolsky}},
  \emph{\bibinfo{title}{Black Holes, White Dwarfs and Neutron Stars}}
  (\bibinfo{publisher}{Wiley}, \bibinfo{address}{New York},
  \bibinfo{year}{1983}).

\bibitem[{\citenamefont{Chandrasekhar}(1939)}]{chan39}
\bibinfo{author}{\bibfnamefont{S.}~\bibnamefont{Chandrasekhar}},
  \emph{\bibinfo{title}{An Introduction to the Study of Stellar Structure}}
  (\bibinfo{publisher}{The University of Chicago Press},
  \bibinfo{address}{Chicago}, \bibinfo{year}{1939}).

\bibitem[{\citenamefont{Artemova et~al.}(1996)\citenamefont{Artemova,
  Bj{\"o}rnsson, and Novikov}}]{art96}
\bibinfo{author}{\bibfnamefont{I.~V.} \bibnamefont{Artemova}},
  \bibinfo{author}{\bibfnamefont{G.}~\bibnamefont{Bj{\"o}rnsson}},
  \bibnamefont{and} \bibinfo{author}{\bibfnamefont{I.~D.}
  \bibnamefont{Novikov}}, \bibinfo{journal}{Astrophys. J.}
  \textbf{\bibinfo{volume}{461}}, \bibinfo{pages}{565} (\bibinfo{year}{1996}).

\bibitem[{\citenamefont{Novikov and Thorne}(1973)}]{nov73}
\bibinfo{author}{\bibfnamefont{I.~D.} \bibnamefont{Novikov}} \bibnamefont{and}
  \bibinfo{author}{\bibfnamefont{K.~S.} \bibnamefont{Thorne}},
  \emph{\bibinfo{title}{Black Holes, ({\rm edited by C. deWitt and B.
  deWitt})}} (\bibinfo{publisher}{Gordon and Breach}, \bibinfo{address}{New
  York}, \bibinfo{year}{1973}).

\bibitem[{\citenamefont{Chakrabarti}(1990)}]{skc90}
\bibinfo{author}{\bibfnamefont{S.~K.} \bibnamefont{Chakrabarti}},
  \emph{\bibinfo{title}{Theory of Transonic Astrophysical Flows}}
  (\bibinfo{publisher}{World Scientific}, \bibinfo{address}{Singapore},
  \bibinfo{year}{1990}).

\bibitem[{\citenamefont{Bond et~al.}(1998)\citenamefont{Bond, Pogosyan, and
  Souradeep}}]{bps1}
\bibinfo{author}{\bibfnamefont{J.~R.} \bibnamefont{Bond}},
  \bibinfo{author}{\bibfnamefont{D.}~\bibnamefont{Pogosyan}}, \bibnamefont{and}
  \bibinfo{author}{\bibfnamefont{T.}~\bibnamefont{Souradeep}},
  \bibinfo{journal}{Class. Quantum Grav.} \textbf{\bibinfo{volume}{15}},
  \bibinfo{pages}{2671} (\bibinfo{year}{1998}).

\bibitem[{\citenamefont{Bond et~al.}(2000)\citenamefont{Bond, Pogosyan, and
  Souradeep}}]{bps2}
\bibinfo{author}{\bibfnamefont{J.~R.} \bibnamefont{Bond}},
  \bibinfo{author}{\bibfnamefont{D.}~\bibnamefont{Pogosyan}}, \bibnamefont{and}
  \bibinfo{author}{\bibfnamefont{T.}~\bibnamefont{Souradeep}},
  \bibinfo{journal}{Phys. Rev. D} \textbf{\bibinfo{volume}{62}},
  \bibinfo{pages}{043005} (\bibinfo{year}{2000}).

\end{thebibliography}

\end{document}